\documentclass[11pt]{article}

\usepackage[preprint]{acl}

\usepackage{times}
\usepackage{latexsym}
\usepackage[T1]{fontenc}
\usepackage[utf8]{inputenc}
\usepackage{microtype}
\usepackage{inconsolata}

\usepackage{amsmath}

\usepackage{url}
\usepackage{graphicx}
\usepackage{booktabs}
\usepackage{float}
\usepackage{amssymb}
\usepackage{fvextra}
\usepackage{newunicodechar}

\newunicodechar{↔}{\ensuremath{\leftrightarrow}}
\newunicodechar{→}{\ensuremath{\rightarrow}}
\newunicodechar{…}{\ldots}
\DefineVerbatimEnvironment{promptbox}{Verbatim}{%
  frame=single, framesep=2mm, fontsize=\scriptsize,
  breaklines, breaksymbolleft={\tiny\ensuremath{\hookrightarrow}}}

\floatstyle{ruled}
\newfloat{algorithm}{t}{lop}
\floatname{algorithm}{Algorithm}

\newcounter{algline}
\newcommand{\algreset}{\setcounter{algline}{0}}
\newcommand{\algline}[2][0]{%
  \stepcounter{algline}%
  \noindent\makebox[1.6em][r]{\footnotesize\thealgline:}%
  \hspace{#1em}#2\\%
}
\newcommand{\algkw}[1]{\textbf{#1}}

\newtheorem{definition}{Definition}

\newcommand{\dataset}{SCRBench}
\newcommand{\pbase}{Base}
\newcommand{\pcot}{Base+CoT}
\newcommand{\pshot}{Base+one-shot}
\newcommand{\pcotshot}{Base+CoT+one-shot}

\title{\dataset: Beyond Equivalence Checking for Semantic Change Reasoning in Code}

\author{
  Omid Saghatchian \\
  {\small Department of Computer and Data Sciences} \\
  {\small Faculty of Mathematical Sciences} \\
  {\small Shahid Beheshti University} \\
  {\small Tehran, Iran} \\
  \texttt{\small o.saghatchian@mail.sbu.ac.ir}
  \And
  Ali Mahzoun \\
  {\small Independent Researcher} \\
  \texttt{\small ali@mahzoun.me}
  \AND
  Khashayar Etemadi \\
  {\small Department of Computer Science} \\
  {\small ETH Zurich} \\
  {\small Zurich, Switzerland} \\
  \texttt{\small ketemadi@ethz.ch}
  \And
  Saeed Reza Kheradpisheh \\
  {\small Department of Computer and Data Sciences} \\
  {\small Faculty of Mathematical Sciences} \\
  {\small Shahid Beheshti University} \\
  {\small Tehran, Iran} \\
  \texttt{\small s\_kheradpisheh@sbu.ac.ir}
}

\begin{document}

\maketitle

\begin{abstract}
Understanding how code changes affect program semantics is a core aspect of code reasoning in language models.
It is commonly evaluated by asking whether two programs are semantically equivalent, but a yes/no answer does not reveal whether a model knows which code change caused a semantic difference, or which sequence of code changes produced the modified program.
We study two tasks that ask these questions directly, diagnosing semantic change and reconstructing sequences of code changes, alongside detecting semantic equivalence.
We build \dataset{}, 67{,}658 Python program pairs generated by applying sequences of code changes that preserve or alter program semantics, and evaluate five language models on a stratified subset of 4{,}500 pairs.
Detecting semantic equivalence exposes weaker models, whose accuracy declines as changes accumulate, but the strongest models remain near the ceiling.
Diagnosing semantic change reveals weaknesses that equivalence accuracy hides: a model with 94.71\% equivalence accuracy diagnoses only 54.44\% of semantic changes.
Reconstructing sequences of code changes is much harder: the best model reconstructs the exact sequence for only 39.27\% of pairs, and no model exceeds 1.11\% when five changes are applied.
High equivalence accuracy therefore overestimates how well models reason about the semantic effects of code changes.
\end{abstract}

\section{Introduction}
\label{sec:intro}
Understanding how code changes affect program semantics is a core aspect of code reasoning in large language models (LLMs).
Evaluating this understanding has received increasing attention \citep{wei2025equibench}.
One common test asks a model whether two programs are semantically equivalent, that is, whether they produce the same output for every input \citep{oza2025cetbench}.
Existing benchmarks commonly construct such program pairs from seed programs by applying small automated code changes, called transformations, that either preserve or alter a program's semantics \citep{wei2025equibench,maveli2025seqcobench,oza2025cetbench}.
In the simplest case, only a single transformation is applied \citep{maveli2025seqcobench}, so the evaluation does not test a model as transformations accumulate. \textsc{CETBench} addresses this by composing several transformations into one transformation sequence \citep{oza2025cetbench}, but it still asks only whether the two programs are semantically equivalent.
A yes/no answer, however, does not reveal how the model arrived at it. A model may correctly judge two programs as non-equivalent without knowing which transformation is responsible for the semantic difference or which sequence of transformations produced the modified program from the original. Equivalence accuracy alone therefore cannot measure this finer-grained reasoning. Measuring it requires asking the model these questions directly.

In this paper, we go a step further: beyond asking whether two programs are semantically equivalent, we ask which transformation changed their semantics and in which order the transformations were applied.
To this end, we define three tasks on pairs of an original program and a modified program produced by a sequence of transformations.
(T1)~\emph{Detecting semantic equivalence}: decide whether the two programs are semantically equivalent.
(T2)~\emph{Diagnosing semantic change}: given a non-equivalent pair, identify which transformation altered the semantics.
(T3)~\emph{Reconstructing transformation sequences}: recover the ordered sequence of transformations that produces the modified program from the original one.
To investigate LLM performance on these tasks, we build \dataset{} from 418 MBPP programs \citep{austin2021program}.
We use each MBPP program as an original program and apply to it a sequence of one to five transformations from a catalog of nine semantics-preserving and semantics-altering transformations (Table~\ref{tab:transformations}).
In every non-equivalent pair, exactly one of the applied transformations alters semantics, so T2 has a single correct answer.
Every pair carries ground truth for all three tasks (Section~\ref{sec:datagen}).
\dataset{} contains 67{,}658 program pairs.
From it we draw a stratified subset of 4{,}500 pairs to evaluate all three tasks as transformations accumulate: for each number of applied transformations from one to five, it contains 450 equivalent and 450 non-equivalent pairs.
T1 and T3 use all 4{,}500 pairs, and T2, which applies only to non-equivalent pairs, uses the 2{,}250 non-equivalent ones.

We evaluate five language models on the stratified subset with a prompt that includes chain-of-thought reasoning and a one-shot example.
Within each task, we analyze how accuracy changes as transformations accumulate (Sections~\ref{sec:exp:rq1}--\ref{sec:exp:rq3}).
We also perform an ablation study on a single model to assess the impact of chain-of-thought reasoning and the one-shot example on accuracy (Sections~\ref{sec:exp:rq1}--\ref{sec:exp:rq3}).
Finally, we analyze the results across the three tasks to characterize the progression of code reasoning, including the opportunities it reveals for improving advanced models (Section~\ref{sec:exp:across}).
Detecting semantic equivalence (T1) and diagnosing semantic change (T2) show substantial variation across models, and T2 reveals differences that T1 hides.
For example, Mistral-Small-24B is within five percentage points of the strongest models on T1 (94.71\%) but about 40 points behind them on T2 (54.44\%).
Across all models, T1 accuracy ranges from 99.69\% for Qwen3-235B to 55.49\% for GPT-4.1-nano, and the strongest models remain accurate as transformations accumulate, while weaker models decline.
T2 accuracy ranges from 97.07\% for GPT-OSS-120B to 22.67\% for Llama-3.1-8B.
Reconstructing transformation sequences (T3) is more demanding, because it requires recovering the ordered sequence of transformations.
Even models that are accurate on T1 and T2 show substantially lower accuracy on T3.
GPT-OSS-120B, the strongest model on T3, reconstructs 39.27\% of the sequences, and no model exceeds 1.11\% when five transformations are applied.
Qwen3-235B, the most accurate model on T1 (99.69\%), reconstructs only 26.82\% of the sequences.

Our contributions are twofold:
(i)~We study two tasks beyond detecting semantic equivalence (T1), diagnosing semantic change (T2) and reconstructing transformation sequences (T3), and show with five language models that T1 accuracy can overstate how well a model understands the effect of code changes on program semantics: T1 exposes weaker models, T2 separates models that perform similarly on T1, and models near the ceiling on T1 and T2 still struggle on T3 (Section~\ref{sec:experiments}).
(ii)~To support this study, we build \dataset{}, 67{,}658 program pairs with automatically checked ground truth for all three tasks: an equivalence label, the single semantics-altering transformation of each non-equivalent pair, and every transformation sequence connecting the two programs, so that any valid order counts as correct (Section~\ref{sec:datagen}); we will publicly release \dataset{} and our code upon conference acceptance.

\begin{table*}[t]
\caption{The nine transformations in the catalog $\mathcal{T}$.}
\label{tab:transformations}
\begin{center}
\small
\begin{tabular}{@{}lllp{10cm}@{}}
\toprule
\textbf{ID} & \textbf{Abbr.} & \textbf{Type} & \textbf{Effect} \\
\midrule
1 & AOM & Altering    & Swaps $+{\leftrightarrow}-$ and $*{\leftrightarrow}/$. \\
2 & BLM & Altering    & Negates the first \texttt{True} (else the first \texttt{False}). \\
3 & LOM & Altering    & Swaps \texttt{and}${\leftrightarrow}$\texttt{or}. \\
4 & COM & Altering    & Flips each comparison's leading operator ($<{\leftrightarrow}>$, $\leq{\leftrightarrow}\geq$, $={\leftrightarrow}\neq$). \\
\addlinespace
5 & RV      & Preserving & Renames the most frequent variable to \texttt{VAR\_0}. \\
6 & DCI\_V1 & Preserving & Inserts an unused assignment. \\
7 & DCI\_V2 & Preserving & Inserts a never-executed \texttt{if} block. \\
8 & OS      & Preserving & Swaps the operands of a comparison between two variables ($a>b \Rightarrow b<a$). \\
9 & LT      & Preserving & Rewrites a \texttt{for}-\texttt{range} loop as a \texttt{while} loop. \\
\bottomrule
\end{tabular}
\end{center}
\end{table*}

\section{Data Generation}
\label{sec:datagen}

We build \dataset, a dataset of program pairs derived from MBPP programs.
Each pair is represented as $(p, p')$, where $p$ is an original program from MBPP and
$p'$ is the modified program obtained by applying a sequence of one to five deterministic transformations to $p$.
Each pair is annotated with a label $y$ indicating whether the two programs are semantically equivalent and with the set of all valid transformation sequences that produce $p'$ from $p$.
Section~\ref{sec:datagen:base} covers the preprocessing of MBPP into a base corpus of original programs.
Section~\ref{sec:datagen:transformations} introduces the transformations and defines valid transformation sequences and, for each pair, the transformation sequence set and length (Definitions~\ref{def:valid-transformation-sequence}--\ref{def:transformation-sequence-set}).
Section~\ref{sec:datagen:generate} describes how the transformations are applied to generate the modified programs.
Section~\ref{sec:datagen:corpus} describes how the generated sequences form labeled pairs and reports the resulting corpus and the stratified subset drawn from it.

\subsection{Base corpus}
\label{sec:datagen:base}

MBPP \citep{austin2021program} is a crowd-sourced dataset of short, introductory Python programs collected for program synthesis, each paired with assert-based unit tests.
We pool the train, validation, and test splits of the sanitized, hand-verified subset of MBPP (420 programs) and preprocess them in three steps.
First, we canonicalize each program by parsing it to an abstract syntax tree (AST) and unparsing it.
The transformations also produce code in this canonical form, so an original program and its modified program differ only in the applied transformations, not in formatting or comments.
Second, we add one hand-written edge-case assertion to each of eight programs whose test suites lacked one, so that the tests also catch semantic changes that only affect edge-case inputs.
Third, we remove the two programs whose canonical form duplicates another program.
This yields a base corpus of 418 programs.
Each entry comprises an original program $p$ and its test suite $E$.

\subsection{Transformations}
\label{sec:datagen:transformations}

A \emph{transformation} $\tau$ is a deterministic function that rewrites a program.
It is \emph{semantics-altering} if it is designed to change the program's semantics and \emph{semantics-preserving} otherwise.
Table~\ref{tab:transformations} lists the nine transformations we use, adapted from \textsc{SeqCoBench} \citep{maveli2025seqcobench}: four altering ($\mathcal{A}$, ids~1--4) and five preserving ($\mathcal{P}$, ids~5--9), which together form the catalog $\mathcal{T} = \mathcal{A} \cup \mathcal{P}$.
Appendix~\ref{app:transformations-choice} motivates this choice, and Appendix~\ref{app:transformations} gives an example of each.

To let transformations accumulate, we apply them in sequence.
Applying a transformation sequence $(\tau_1,\dots,\tau_k)$ to an original program $p = c_0$ produces the programs $c_i = \tau_i(c_{i-1})$ and ends in the modified program $p' = c_k$.
We keep only sequences that meet three conditions, formalized in Definition~\ref{def:valid-transformation-sequence}.
First, each transformation appears at most once, because applying a transformation twice can undo it.
For example, a second AOM swaps every $-$ back to $+$.
Second, at most one transformation is altering, so that T2 has a single correct answer.
Third, every transformation has its intended effect, and the tests confirm it.
A transformation can leave the program unchanged, for example AOM on a program without arithmetic operators.
An altering transformation can also change only code that the tests never execute, so the tests cannot confirm the semantic change.
A transformation can also make a program run forever, for example when AOM turns a loop increment into a decrement, so we also reject a transformation if the tests on its output exceed a time limit.

\begin{definition}[Valid transformation sequence]
\label{def:valid-transformation-sequence}
A transformation sequence $(\tau_1,\dots,\tau_k)$ applied to $c_0$ with tests $E$ is \emph{valid} iff
(1)~its transformation ids are pairwise distinct,
(2)~at most one $\tau_i$ is altering, and
(3)~every $\tau_i$ is \emph{valid}: it changes the AST of $c_{i-1}$, the tests on $c_i$ finish within a time limit, and $c_i$ passes all of them if $\tau_1,\dots,\tau_i$ are all preserving and fails at least one otherwise.
\end{definition}

Different valid sequences can produce the same pair, for example the same transformations applied in a different order.
We therefore annotate each pair with all of them.
Appendix~\ref{app:example-pair} shows a pair with two such sequences.
The length of these sequences is the number of transformations that separate $p$ from $p'$.
We use it to stratify the subset and to report how accuracy changes as transformations accumulate.

\begin{definition}[Sequence set and length]
\label{def:transformation-sequence-set}
For a pair $(p, p')$, the \emph{transformation sequence set} $\Pi(p, p')$ is the set of all valid transformation sequences of length at most five that produce $p'$ from $p$.
All sequences in $\Pi(p, p')$ have the same length, which Appendix~\ref{app:aggregate} verifies for every pair in \dataset{}; this length is the \emph{transformation sequence length} $k(p, p')$.
\end{definition}

\subsection{Incremental generation}
\label{sec:datagen:generate}

We now generate program pairs from the base corpus.
The first $i$ transformations of a sequence form a prefix, which produces $c_i$.
We group prefixes into collections $D_{k,a}$ by their length $k$ and by the position $a$ of their altering transformation, with $a=0$ if they contain none.
The base corpus is $D_{0,0}$.
Algorithm~\ref{alg:collections} builds each collection by extending the prefixes of a parent collection with transformations from a pool, either $\mathcal{A}$ or $\mathcal{P}$.

\begin{algorithm}[t]
\caption{\textsc{BuildCollections} --- construct every $D_{k,a}$ from its parent collection.}
\label{alg:collections}
\algreset
\algline{\algkw{Input:} base corpus $D_{0,0}$}
\algline{\algkw{Output:} collections $D_{k,a}$ for $1 \le k \le 5$, $0 \le a \le k$}
\algline{\algkw{for} $k = 1$ to $5$ \algkw{do}}
\algline[1]{\algkw{for} $a = 0$ to $k$ \algkw{do}}
\algline[2]{\algkw{if} $a = 0$ \algkw{then}}
\algline[3]{$D_{\mathrm{parent}} \gets D_{k-1,\,0}$;\quad $\mathcal{T}_{\mathrm{pool}} \gets \mathcal{P}$}
\algline[2]{\algkw{else if} $a = k$ \algkw{then}}
\algline[3]{$D_{\mathrm{parent}} \gets D_{k-1,\,0}$;\quad $\mathcal{T}_{\mathrm{pool}} \gets \mathcal{A}$}
\algline[2]{\algkw{else}}
\algline[3]{$D_{\mathrm{parent}} \gets D_{k-1,\,a}$;\quad $\mathcal{T}_{\mathrm{pool}} \gets \mathcal{P}$}
\algline[2]{$D_{k,a} \gets \textsc{Extend}(D_{\mathrm{parent}},\, \mathcal{T}_{\mathrm{pool}})$}
\algline{\algkw{return} $\{D_{k,a}\}_{1 \le k \le 5,\, 0 \le a \le k}$}
\end{algorithm}

The algorithm distinguishes three cases.
If $a=0$, it appends a preserving transformation to a prefix of $D_{k-1,0}$, which contains no altering transformation.
If $a=k$, it appends an altering transformation to a prefix of $D_{k-1,0}$.
If $1 \le a < k$, it appends a preserving transformation to a prefix of $D_{k-1,a}$, which already contains its altering transformation.
Every prefix therefore contains at most one altering transformation.
$\textsc{Extend}$ appends each transformation of the pool that is not yet in the prefix and marks the new prefix as valid or invalid by Definition~\ref{def:valid-transformation-sequence}.
It extends only valid prefixes, because a sequence with an invalid prefix is itself invalid, so no valid sequence is lost.
Appendix~\ref{app:algorithm} gives the full procedure.

\subsection{Corpus}
\label{sec:datagen:corpus}

Each valid prefix of length $k$ is a valid transformation sequence from $p$ to a program $c_k$.
We group these sequences by the pair $(p, p') = (p, c_k)$ that they produce and record each pair with its transformation sequence set $\Pi(p, p')$.
The pair is labeled non-equivalent ($y=0$) if a sequence in $\Pi(p, p')$ contains an altering transformation, and equivalent ($y=1$) otherwise.
Appendix~\ref{app:aggregate} also verifies that either all sequences in $\Pi(p, p')$ contain an altering transformation or none does, so $y$ is well-defined.

From the 418 original programs, we obtain 67{,}658 labeled pairs, of which 12{,}510 are equivalent and 55{,}148 are non-equivalent.
Table~\ref{tab:stats} in Appendix~\ref{app:corpus-stats} breaks them down by transformation sequence length $k$.
The pairs are unevenly distributed: most are non-equivalent, and 89\% have $k \ge 3$.
To weigh every length and label equally, we use stratified sampling: we split the pairs into ten strata, one for each combination of $k$ and label, and draw the same number of pairs from each stratum.
The smallest stratum, the non-equivalent pairs at $k{=}1$, limits this number, and we draw 450 pairs per stratum.
The resulting stratified subset contains 4{,}500 pairs, 2{,}250 equivalent and 2{,}250 non-equivalent.
Within each stratum, we choose the pairs so that even the least frequent transformation appears in as many pairs as possible (Appendix~\ref{app:sampling}).

\section{Experiments}
\label{sec:experiments}

We evaluate five language models on the three tasks defined in Section~\ref{sec:intro}, using labeled program pairs $(p, p')$ from \dataset{}.
Section~\ref{sec:exp:setup} describes the common experimental setup.
Sections~\ref{sec:exp:rq1}--\ref{sec:exp:rq3} report the results for T1, T2, and T3, respectively, and Section~\ref{sec:exp:across} examines the progression of code reasoning capabilities across tasks.

\subsection{Experiment Setup}
\label{sec:exp:setup}

For each task, we first compare four prompt variants on a single model, Qwen3-235B, and then compare five models using a prompt with chain-of-thought reasoning and a one-shot example.
In both, we report overall accuracy and accuracy by transformation sequence length $k$ (Section~\ref{sec:datagen}).

\paragraph{Models.}
We evaluate five language models:
Qwen3-235B~\citep{yang2025qwen3}, an open-weight mixture-of-experts model with 22B active parameters;
GPT-OSS-120B~\citep{openai2025gptoss}, an open-weight mixture-of-experts reasoning model;
Mistral-Small-24B~\citep{mistral2025small3} and Llama-3.1-8B~\citep{grattafiori2024llama3}, two open-weight dense models;
and GPT-4.1-nano~\citep{openai2025gpt41}, a proprietary model of undisclosed size.
Appendix~\ref{app:reproducibility} lists the exact model identifiers.

\paragraph{Prompting.}
We construct four prompt variants by independently adding chain-of-thought (CoT) reasoning and a one-shot example to the base instruction: \pbase{}, \pcot{}, \pshot{}, and \pcotshot{} (Table~\ref{tab:prompt-variants}; templates in Appendices~\ref{app:prompts:rq1}--\ref{app:prompts:rq3}).

\paragraph{Model execution and response processing.}
We perform inference at temperature $0$ with a maximum generation budget of $4{,}000$ tokens and obtain one response for each program pair.
A rule-based parser extracts the model's answer from each response.
Responses that match no parsing rule are classified as \emph{unparseable} and counted as incorrect.

\subsection{T1: Detecting Semantic Equivalence}
\label{sec:exp:rq1}

In T1, the model receives an original program $p$ and a modified program $p'$ and predicts whether they are semantically equivalent.
The ground-truth label is $y$ from Section~\ref{sec:datagen:corpus}.
The parser extracts either \texttt{YES} or \texttt{NO} from the response.
We evaluate T1 on all 4{,}500 pairs of the stratified subset.

Table~\ref{tab:rq1} reports T1 accuracy.
For Qwen3, prompt choice substantially affects T1 accuracy: \pcotshot{} performs best and \pshot{} worst, with a 13.96-percentage-point gap.
Adding CoT raises T1 accuracy by 11.58 percentage points without the one-shot example (from 88.02\% to 99.60\%) and by 13.96 percentage points with it (from 85.73\% to 99.69\%).
Once CoT is included, the one-shot example has almost no effect, improving accuracy by only 0.09 percentage points; without CoT, it lowers accuracy by 2.29 percentage points.
These results indicate that adding the CoT instruction substantially improves T1 performance.

\begin{table*}[t]
\caption{T1 accuracy across prompt and model variants.}
\label{tab:rq1}
\begin{center}
\small
\begin{tabular}{@{}llrrrrrr@{}}
\toprule
\textbf{Model} & \textbf{Prompt} & \textbf{Overall} & $k{=}1$ & $k{=}2$ & $k{=}3$ & $k{=}4$ & $k{=}5$ \\
\midrule
\textbf{Qwen3-235B} & \pbase{} & 88.02\% & 98.33\% & 91.78\% & 88.11\% & 82.56\% & 79.33\% \\
\textbf{Qwen3-235B} & \pcot{} & 99.60\% & 100.00\% & 99.56\% & 99.56\% & 99.56\% & 99.33\% \\
\textbf{Qwen3-235B} & \pshot{} & 85.73\% & 97.44\% & 89.67\% & 87.56\% & 78.89\% & 75.11\% \\
\textbf{Qwen3-235B} & \pcotshot{} & 99.69\% & 99.89\% & 99.22\% & 99.56\% & 99.89\% & 99.89\% \\
\midrule
\textbf{GPT-OSS-120B} & \pcotshot{} & 98.16\% & 99.22\% & 98.56\% & 98.67\% & 97.22\% & 97.11\% \\
\textbf{Mistral-Small-24B} & \pcotshot{} & 94.71\% & 98.89\% & 95.56\% & 93.56\% & 93.67\% & 91.89\% \\
\textbf{Llama-3.1-8B} & \pcotshot{} & 67.93\% & 78.56\% & 70.78\% & 67.44\% & 62.22\% & 60.67\% \\
\textbf{GPT-4.1-nano} & \pcotshot{} & 55.49\% & 58.00\% & 55.56\% & 54.00\% & 56.22\% & 53.67\% \\
\bottomrule
\end{tabular}
\end{center}
\end{table*}

Under the designated \pcotshot{} prompt, accuracy is generally highest at $k=1$ and tends to decrease as $k$ increases.
Qwen3 remains near ceiling, varying by at most 0.67 percentage points across sequence lengths, whereas Llama-3.1 declines by 17.89 percentage points, from 78.56\% to 60.67\%.
This pattern indicates that longer transformation sequences are associated with lower semantic-equivalence accuracy, particularly for weaker models.
This pattern is consistent with \citet{laneve2025assessing}, who find that combining two semantics-preserving transformations lowers equivalence accuracy below that of either transformation alone.
T1 therefore measures not only whether models can detect semantic equivalence, but also how robustly they maintain that judgment as transformation sequences become longer.
Performance on single-transformation pairs can overstate a model's ability to reason about equivalence under more complex transformation sequences.

For some weaker models, overall accuracy can also obscure class-specific bias.
For example, GPT-4.1-nano correctly classifies only 11.64\% of equivalent pairs but 99.33\% of non-equivalent pairs, indicating that its aggregate performance is largely driven by predicting \texttt{NO}.
Because the one-shot example also has a \texttt{NO} answer, we additionally evaluate this model with the CoT prompt without the one-shot example to examine whether the example contributes to this bias.
Removing the example raises overall accuracy from 55.49\% to 64.69\%, but the class-specific accuracies remain imbalanced at 31.16\% and 98.22\% for equivalent and non-equivalent pairs, respectively.

Taken together, these results expose two limitations of weaker models on T1: accuracy can decline as transformation sequences grow, and aggregate accuracy can be driven by class-biased predictions.
For stronger models, the limitation lies in T1 itself: near-ceiling accuracy does not by itself establish that they can reason about the semantic effects of individual transformations, because T1 requires only a binary equivalence judgment; T2 addresses this by asking which transformation altered the semantics.

\subsection{T2: Diagnosing Semantic Change}
\label{sec:exp:rq2}

Given a non-equivalent program pair $(p,p')$ and the transformation catalog $\mathcal{T}$ (see Appendix~\ref{app:prompts:catalog}), T2 requires the model to identify which transformation is responsible for the semantic difference between the programs.
We evaluate T2 on the 2{,}250 non-equivalent pairs of the stratified subset.

Table~\ref{tab:rq2} reports T2 accuracy.
CoT substantially improves T2 accuracy: on Qwen3, it raises accuracy from 43.51\% to 93.51\% without the one-shot example and from 37.29\% to 93.47\% with it.
The one-shot example does not improve T2 accuracy: it lowers accuracy from 43.51\% to 37.29\% without CoT and from 93.51\% to 93.47\% with CoT.

\begin{table*}[t]
\caption{T2 accuracy across prompt and model variants.}
\label{tab:rq2}
\begin{center}
\small
\begin{tabular}{@{}llrrrrrr@{}}
\toprule
\textbf{Model} & \textbf{Prompt} & \textbf{Overall} & $k{=}1$ & $k{=}2$ & $k{=}3$ & $k{=}4$ & $k{=}5$ \\
\midrule
\textbf{Qwen3-235B} & \pbase{} & 43.51\% & 93.78\% & 52.44\% & 29.11\% & 24.00\% & 18.22\% \\
\textbf{Qwen3-235B} & \pcot{} & 93.51\% & 100.00\% & 92.67\% & 92.22\% & 91.33\% & 91.33\% \\
\textbf{Qwen3-235B} & \pshot{} & 37.29\% & 96.89\% & 45.56\% & 20.44\% & 15.11\% & 8.44\% \\
\textbf{Qwen3-235B} & \pcotshot{} & 93.47\% & 100.00\% & 93.11\% & 90.67\% & 91.56\% & 92.00\% \\
\midrule
\textbf{GPT-OSS-120B} & \pcotshot{} & 97.07\% & 100.00\% & 95.33\% & 96.22\% & 95.56\% & 98.22\% \\
\textbf{Mistral-Small-24B} & \pcotshot{} & 54.44\% & 94.00\% & 58.00\% & 44.00\% & 41.33\% & 34.89\% \\
\textbf{Llama-3.1-8B} & \pcotshot{} & 22.67\% & 70.89\% & 24.67\% & 10.67\% & 3.78\% & 3.33\% \\
\textbf{GPT-4.1-nano} & \pcotshot{} & 61.07\% & 87.78\% & 67.56\% & 56.22\% & 49.56\% & 44.22\% \\
\bottomrule
\end{tabular}
\end{center}
\end{table*}

All models perform best at $k=1$.
T2 accuracy declines modestly for GPT-OSS and Qwen3 but sharply for the other three models, most of all for Llama-3.1 (from 70.89\% to 3.33\%), indicating that diagnosis becomes more difficult as semantics-preserving transformations accumulate around it.

\begin{table}[t]
\caption{T2 accuracy by altering transformation using the designated prompt \pcotshot{}.}
\label{tab:rq2-recall}
\begin{center}
\small
\setlength{\tabcolsep}{4pt}
\begin{tabular}{@{}lrrrr@{}}
\toprule
\textbf{Model} & AOM & BLM & LOM & COM \\
\midrule
Qwen3-235B        & 93.98\% & 95.93\% & 99.77\% & 86.85\% \\
GPT-OSS-120B      & 93.57\% & 99.10\% & 100.00\% & 97.65\% \\
Mistral-Small-24B & 56.77\% & 22.17\% & 62.10\% & 68.86\% \\
Llama-3.1-8B      & 28.73\% &  6.79\% &  6.85\% & 37.56\% \\
GPT-4.1-nano      & 65.25\% & 43.67\% & 51.83\% & 74.65\% \\
\bottomrule
\end{tabular}
\end{center}
\end{table}

Table~\ref{tab:rq2-recall} shows that T2 performance varies across models and types of semantic alteration.
Higher-performing models remain accurate across alteration types, whereas lower-performing models show substantial variation and make different transformation-specific confusions.
For example, Llama-3.1 often predicts preserving transformations such as RV or DCI\_V2, whereas GPT-4.1-nano frequently confuses operator-altering transformations such as AOM and COM.

Because T1 records only a binary equivalence label, it does not reveal which transformation a model attributes the change to; T2 evaluates that attribution directly.
Thus, T2 reveals limitations that equivalence checking alone cannot capture.
However, T2 concerns only the single transformation that altered the semantics; we address this limitation in T3, which evaluates the full transformation sequence.

\subsection{T3: Reconstructing Transformation Sequences}
\label{sec:exp:rq3}

In T3, given $p$, $p'$, and the transformation catalog $\mathcal{T}$, the model predicts an ordered sequence of transformations that maps $p$ to $p'$. A prediction is correct if it is one of the sequences in $\Pi(p, p')$.
Because $k$ is not disclosed in the prompt, the model must infer the sequence length as well as identify and order the transformations.
Reconstructing this transformation history provides a controlled test of change tracking, an ability relevant to software-engineering tasks such as debugging, program repair, and code review.
We evaluate T3 on all 4{,}500 pairs of the stratified subset.

\begin{table*}[t]
\caption{T3 accuracy across prompt and model variants.}
\label{tab:rq3}
\begin{center}
\small
\begin{tabular}{@{}llrrrrrr@{}}
\toprule
\textbf{Model} & \textbf{Prompt} & \textbf{Overall} & $k{=}1$ & $k{=}2$ & $k{=}3$ & $k{=}4$ & $k{=}5$ \\
\midrule
\textbf{Qwen3-235B} & \pbase{} & 25.42\% & 78.89\% & 38.22\% & 7.78\% & 2.11\% & 0.11\% \\
\textbf{Qwen3-235B} & \pcot{} & 27.93\% & 99.33\% & 32.11\% & 7.78\% & 0.33\% & 0.11\% \\
\textbf{Qwen3-235B} & \pshot{} & 22.47\% & 75.44\% & 27.22\% & 8.56\% & 1.11\% & 0.00\% \\
\textbf{Qwen3-235B} & \pcotshot{} & 26.82\% & 99.00\% & 28.56\% & 5.89\% & 0.67\% & 0.00\% \\
\midrule
\textbf{GPT-OSS-120B} & \pcotshot{} & 39.27\% & 99.22\% & 62.56\% & 26.22\% & 7.22\% & 1.11\% \\
\textbf{Mistral-Small-24B} & \pcotshot{} & 25.11\% & 85.11\% & 35.22\% & 4.44\% & 0.78\% & 0.00\% \\
\textbf{Llama-3.1-8B} & \pcotshot{} & 3.09\% & 14.22\% & 1.00\% & 0.22\% & 0.00\% & 0.00\% \\
\textbf{GPT-4.1-nano} & \pcotshot{} & 19.96\% & 71.67\% & 23.44\% & 4.00\% & 0.67\% & 0.00\% \\
\bottomrule
\end{tabular}
\end{center}
\end{table*}

Table~\ref{tab:rq3} reports T3 accuracy.
Prompt variants have limited effect on overall T3 accuracy, indicating that these prompt changes alone do not resolve the difficulty of sequence reconstruction.
Accuracy decreases sharply as $k$ increases for every model.
Because T3 requires the entire sequence to be correct, an error in any transformation, its order, or the inferred length makes the prediction incorrect; as sequences become longer, there are more opportunities for such errors.

The contrast with T1 and T2 is substantial: models that perform strongly at detecting semantic equivalence and diagnosing the semantics-altering transformation still struggle to reconstruct how the modified program was produced.
Thus, success on the preceding tasks does not imply that a model can recover the complete transformation sequence.

\subsection{A Progression of Code Reasoning Capabilities}
\label{sec:exp:across}

\begin{figure}[t]
\centering
\includegraphics[width=\linewidth]{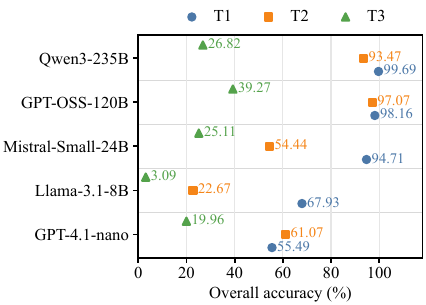}
\caption{Overall accuracy on all tasks for all models, using \pcotshot{}.}
\label{fig:cross-rq}
\end{figure}

The three tasks form a progression from detecting semantic equivalence (T1), the standard task, to the two tasks we introduce: diagnosing semantic change (T2) and reconstructing transformation sequences (T3).
Figure~\ref{fig:cross-rq} presents the models' overall accuracy on each task using \pcotshot{}; each task reveals limitations that the tasks before it do not.

T2 distinguishes models that T1 does not.
On T1, Mistral-Small-24B is within five percentage points of the strongest models (94.71\%), but on T2 it reaches only 54.44\%, while Qwen3-235B and GPT-OSS-120B remain above 93\%.
The ranking also changes: Qwen3 is the most accurate model on T1, whereas GPT-OSS is the most accurate on T2.
High equivalence accuracy therefore does not imply that a model can identify which change altered the semantics.

T3 remains difficult even for models that succeed on both T1 and T2.
Qwen3 and GPT-OSS exceed 93\% on T1 and T2, yet reconstruct only 26.82\% and 39.27\% of sequences, and every model is substantially less accurate on T3 than on either preceding task.
Identifying the altering transformation is thus much easier than recovering the complete ordered sequence.

Together, T2 and T3 complement equivalence checking: T2 exposes differences in attributing semantic change that T1 accuracy hides, and T3 marks a frontier that even the strongest models have not reached, namely tracking every transformation and its order.

\section{Related Work}
\label{sec:related}

There are three lines of research adjacent to our work: (i)~benchmarks that pose the semantic equivalence detection question on transformed program
pairs; (ii)~studies of LLM robustness under semantics-preserving perturbations of code; and (iii)~classical, formal techniques for program equivalence checking.

\paragraph{Benchmarking LLM-based code equivalence detection.}
The prior work closest to \dataset{} builds pair datasets by applying transformations to seed programs and asking a language model whether the two programs are equivalent. \textsc{SeqCoBench} \citep{maveli2025seqcobench} applies a single semantics-preserving or altering transformation to MBPP programs \citep{austin2021program}. \textsc{CETBench} \citep{oza2025cetbench} applies a random sequence of up to five perturbations to pairs of CodeContests solutions and reports a substantial drop in LLM accuracy on the perturbed pairs. \textsc{EquiBench} \citep{wei2025equibench} covers Python, C, CUDA and x86-64 and generates pairs via program analysis, compiler scheduling, and superoptimization.
\citet{allamanis2025probegen} use LLMs with execution feedback to disprove equivalence via counterexample search.
Our work differs from these studies on two axes. First, transformations in \dataset{} are composed sequentially on a single seed, whereas \textsc{SeqCoBench} applies a single transformation and \textsc{CETBench} distributes each perturbation over one of two programs in a starting pair. Second, in addition to detecting semantic equivalence (T1), we evaluate LLMs on diagnosing \emph{which} transformation in the sequence changed the semantics (T2) and on reconstructing the sequence that produces the target from the source (T3). To our knowledge no prior benchmark in this family addresses either.

One of the major applications of semantic equivalence detection techniques is equivalent mutant identification. Mutation testing \citep{jia2011survey,papadakis2019mutation} inserts small syntactic faults into a program to produce \emph{mutants} and measures how many of the mutants are killed by the test suite. An obstacle to mutation testing is the \emph{equivalent mutant problem}: a syntactically distinct mutant may nonetheless be semantically equivalent to the original and therefore unkillable, biasing the mutation score \citep{madeyski2014slr}. Classical equivalent mutant identification techniques include compiler-optimization checks \citep{offutt1994compiler,papadakis2015tce,kintis2018tce}, program slicing \citep{hierons1999slicing}, and constraint and symbolic execution based reasoning \citep{nica2012constraints}; learning-based approaches include feature-engineered classifiers \citep{naeem2020ml}, and AST-based neural models \citep{peacock2021astnn}. More recently LLMs are applied directly to solve this problem \citep{tian2024llm}, typically on benchmarks like \textsc{MutantBench} \citep{vanhijfte2021mutantbench}.

\paragraph{Semantics-preserving perturbations and model robustness.}
A parallel line studies LLM \emph{robustness} on downstream code tasks under
semantics-preserving perturbations of the input. \textsc{ReCode} \citep{wang2023recode} perturbs docstrings, function names, syntax, and format
on HumanEval \citep{chen2021humaneval} and MBPP, whose test suites EvalPlus later strengthened \citep{liu2023evalplus}. \citet{henkel2022semantic} use sequences of semantics-preserving transformations to attack and harden code-summarization models, and \citet{hooda2024counterfactual} apply counterfactual mutations such as branch flipping to probe conceptual understanding. \citet{laneve2025assessing} apply compiler-style transformations such as copy propagation and constant folding and find that models often fail to recognize the rewritten program as still equivalent. Related studies explore LLM code reasoning by asking the model to predict inputs or outputs
of a given program \citep{gu2024cruxeval,liu2024codemind}. All of these works measure how a fixed downstream prediction shifts when the input is rewritten, whereas \dataset{} treats the transformations themselves as first-class objects the model must reason about: whether a chain of transformations alters semantics, which one does, and in what order the transformations are applied.

\paragraph{Formal program equivalence detection.}
Program equivalence is undecidable in general, yet a rich body
of research develops sound techniques for restricted settings: translation validation for optimizing compilers \citep{pnueli1998translation,necula2000translation}, differential symbolic execution \citep{person2008differential}, and semantic program alignment
\citep{churchill2019semantic}. \citet{badihi2021eqbench} released \textsc{EqBench}, a benchmark for evaluating such tools. These formal techniques are typically sound but incomplete and scale poorly with syntactic diversity. The recent LLM-based techniques provide the opportunity for large scale program equivalence analysis, which we study through T1--T3, using tens of thousands of labeled pairs in \dataset{} with various levels of syntactic differences.

\section{Conclusion}
\label{sec:conclusion}

Semantic equivalence checking is a common test of how language models reason about code changes, but a yes/no answer does not reveal whether a model understands the semantic effect of each change.
We therefore studied two tasks beyond (T1) detecting semantic equivalence: (T2) diagnosing semantic change, that is, identifying which transformation altered the semantics, and (T3) reconstructing transformation sequences, that is, recovering the ordered transformations that produced the modified program.
To evaluate language models on these three tasks, we built \dataset{}, a dataset of 67{,}658 Python program pairs generated by applying sequences of semantics-preserving and semantics-altering transformations to 418 MBPP programs, with ground truth for all three tasks; \dataset{} can also support training models on them.
For weaker models, T1 is informative, particularly as transformations accumulate and their accuracy declines.
For stronger models, however, T1 accuracy approaches the ceiling, giving the misleading impression that these models fully understand how code changes affect program semantics.
The two new tasks expose what T1 hides: T2 separates models that perform similarly on T1, and T3 remains difficult even for the strongest models, the best of which reconstructs only 39.27\% of sequences.
Semantic equivalence checking is therefore a useful but incomplete measure of code reasoning.
T2 and T3 show where even strong models can still improve: attributing and tracking code changes, as needed in debugging, program repair, and code review.

\section*{Limitations}
\dataset{} consists of short Python functions from MBPP and a catalog of nine transformations adapted from \textsc{SeqCoBench}, so it does not cover larger programs, other languages, or more complex transformations.
The equivalence labels also depend on tests: a pair is labeled equivalent if every applied transformation is designed to preserve semantics and the MBPP tests still pass, which does not prove equivalence.
The non-equivalent labels are certain, because a failing test shows that the two programs behave differently.
The cross-model comparison uses a single prompt variant with one response per pair at temperature $0$.
Extending the program sources and the transformation catalog, and evaluating additional models and prompting strategies, are directions for future work.

\section*{Ethics Statement}
\dataset{} is derived from MBPP \citep{austin2021program}, which is released under the CC~BY~4.0 license \citep{mbpp2021dataset}, and contains only short, self-contained Python programming problems written for benchmarking.
It contains no personal data, and our study involves no human subjects.
All models are accessed through public APIs under their providers' terms of use.
We do not foresee direct negative societal impact; the benchmark is intended to measure, and ultimately improve, how reliably language models reason about the semantic effect of code changes.

\bibliography{references}

\begin{thebibliography}{37}
\providecommand{\natexlab}[1]{#1}

\bibitem[{Allamanis and Yin(2025)}]{allamanis2025probegen}
Miltiadis Allamanis and Pengcheng Yin. 2025.
\newblock \href {https://doi.org/10.48550/arXiv.2502.18473} {Disproving program
  equivalence with {LLMs}}.
\newblock \emph{arXiv preprint arXiv:2502.18473}.

\bibitem[{Austin et~al.(2021)Austin, Odena, Nye, Bosma, Michalewski, Dohan,
  Jiang, Cai, Terry, Le, and Sutton}]{austin2021program}
Jacob Austin, Augustus Odena, Maxwell Nye, Maarten Bosma, Henryk Michalewski,
  David Dohan, Ellen Jiang, Carrie Cai, Michael Terry, Quoc Le, and Charles
  Sutton. 2021.
\newblock \href {https://doi.org/10.48550/arXiv.2108.07732} {Program synthesis
  with large language models}.
\newblock \emph{arXiv preprint arXiv:2108.07732}.

\bibitem[{Badihi et~al.(2021)Badihi, Li, and Rubin}]{badihi2021eqbench}
Sahar Badihi, Yi~Li, and Julia Rubin. 2021.
\newblock \href {https://doi.org/10.1109/MSR52588.2021.00084} {{EqBench}: A
  dataset of equivalent and non-equivalent program pairs}.
\newblock In \emph{2021 IEEE/ACM 18th International Conference on Mining
  Software Repositories (MSR)}, pages 610--614. IEEE.

\bibitem[{Chakraborty et~al.(2022)Chakraborty, Ahmed, Ding, Devanbu, and
  Ray}]{chakraborty2022natgen}
Saikat Chakraborty, Toufique Ahmed, Yangruibo Ding, Premkumar~T. Devanbu, and
  Baishakhi Ray. 2022.
\newblock \href {https://doi.org/10.1145/3540250.3549162} {{NatGen}: Generative
  pre-training by ``naturalizing'' source code}.
\newblock In \emph{Proceedings of the 30th {ACM} Joint European Software
  Engineering Conference and Symposium on the Foundations of Software
  Engineering}, pages 18--30.

\bibitem[{Chen et~al.(2021)Chen, Tworek, Jun, Yuan, Pinto, Kaplan, Edwards,
  Burda, Joseph, Brockman et~al.}]{chen2021humaneval}
Mark Chen, Jerry Tworek, Heewoo Jun, Qiming Yuan, Henrique Ponde De~Oliveira
  Pinto, Jared Kaplan, Harri Edwards, Yuri Burda, Nicholas Joseph, Greg
  Brockman, et~al. 2021.
\newblock \href {https://doi.org/10.48550/arXiv.2107.03374} {Evaluating large
  language models trained on code}.
\newblock \emph{arXiv preprint arXiv:2107.03374}.

\bibitem[{Churchill et~al.(2019)Churchill, Padon, Sharma, and
  Aiken}]{churchill2019semantic}
Berkeley Churchill, Oded Padon, Rahul Sharma, and Alex Aiken. 2019.
\newblock \href {https://doi.org/10.1145/3314221.3314596} {Semantic program
  alignment for equivalence checking}.
\newblock In \emph{Proceedings of the 40th ACM SIGPLAN Conference on
  Programming Language Design and Implementation}, pages 1027--1040.

\bibitem[{{Google Research}(2021)}]{mbpp2021dataset}
{Google Research}. 2021.
\newblock \href {https://huggingface.co/datasets/google-research-datasets/mbpp}
  {{MBPP}: {Mostly Basic Python Problems}}.
\newblock \url{https://huggingface.co/datasets/google-research-datasets/mbpp}.
\newblock Dataset card; released under the CC BY 4.0 license.

\bibitem[{Grattafiori et~al.(2024)Grattafiori, Dubey, Jauhri, Pandey, Kadian
  et~al.}]{grattafiori2024llama3}
Aaron Grattafiori, Abhimanyu Dubey, Abhinav Jauhri, Abhinav Pandey, Abhishek
  Kadian, et~al. 2024.
\newblock \href {https://doi.org/10.48550/arXiv.2407.21783} {The {Llama} 3 herd
  of models}.
\newblock \emph{arXiv preprint arXiv:2407.21783}.

\bibitem[{Gu et~al.(2024)Gu, Rozi{\`e}re, Leather, Solar-Lezama, Synnaeve, and
  Wang}]{gu2024cruxeval}
Alex Gu, Baptiste Rozi{\`e}re, Hugh~James Leather, Armando Solar-Lezama,
  Gabriel Synnaeve, and Sida~I. Wang. 2024.
\newblock \href {https://proceedings.mlr.press/v235/gu24c.html} {{CRUXEval}: A
  benchmark for code reasoning, understanding and execution}.
\newblock In \emph{Proceedings of the 41st International Conference on Machine
  Learning}, volume 235 of \emph{Proceedings of Machine Learning Research},
  pages 16568--16621. PMLR.

\bibitem[{Henkel et~al.(2022)Henkel, Ramakrishnan, Wang, Albarghouthi, Jha, and
  Reps}]{henkel2022semantic}
Jordan Henkel, Goutham Ramakrishnan, Zi~Wang, Aws Albarghouthi, Somesh Jha, and
  Thomas Reps. 2022.
\newblock \href {https://doi.org/10.1109/SANER53432.2022.00070} {Semantic
  robustness of models of source code}.
\newblock In \emph{2022 IEEE International Conference on Software Analysis,
  Evolution and Reengineering (SANER)}, pages 526--537. IEEE.

\bibitem[{Hierons et~al.(1999)Hierons, Harman, and
  Danicic}]{hierons1999slicing}
Rob Hierons, Mark Harman, and Sebastian Danicic. 1999.
\newblock \href
  {https://doi.org/10.1002/(SICI)1099-1689(199912)9:4<233::AID-STVR191>3.0.CO;2-3}
  {Using program slicing to assist in the detection of equivalent mutants}.
\newblock \emph{Software Testing, Verification and Reliability}, 9(4):233--262.

\bibitem[{Hooda et~al.(2024)Hooda, Christodorescu, Allamanis, Wilson, Fawaz,
  and Jha}]{hooda2024counterfactual}
Ashish Hooda, Mihai Christodorescu, Miltiadis Allamanis, Aaron Wilson, Kassem
  Fawaz, and Somesh Jha. 2024.
\newblock \href {https://proceedings.mlr.press/v235/hooda24a.html} {Do large
  code models understand programming concepts? {C}ounterfactual analysis for
  code predicates}.
\newblock In \emph{Proceedings of the 41st International Conference on Machine
  Learning}, volume 235 of \emph{Proceedings of Machine Learning Research},
  pages 18738--18748. PMLR.

\bibitem[{Jia and Harman(2011)}]{jia2011survey}
Yue Jia and Mark Harman. 2011.
\newblock \href {https://doi.org/10.1109/TSE.2010.62} {An analysis and survey
  of the development of mutation testing}.
\newblock \emph{IEEE Transactions on Software Engineering}, 37(5):649--678.

\bibitem[{Kintis et~al.(2018)Kintis, Papadakis, Jia, Malevris, Le~Traon, and
  Harman}]{kintis2018tce}
Marinos Kintis, Mike Papadakis, Yue Jia, Nicos Malevris, Yves Le~Traon, and
  Mark Harman. 2018.
\newblock \href {https://doi.org/10.1109/TSE.2017.2684805} {Detecting trivial
  mutant equivalences via compiler optimisations}.
\newblock \emph{IEEE Transactions on Software Engineering}, 44(4):308--333.

\bibitem[{Laneve et~al.(2025)Laneve, Span{\`o}, Ressi, Rossi, and
  Bugliesi}]{laneve2025assessing}
Cosimo Laneve, Alvise Span{\`o}, Dalila Ressi, Sabina Rossi, and Michele
  Bugliesi. 2025.
\newblock \href {https://doi.org/10.1007/978-3-031-95497-9\_13} {Assessing code
  understanding in {LLM}s}.
\newblock In \emph{Formal Techniques for Distributed Objects, Components, and
  Systems (FORTE 2025)}, volume 15732 of \emph{Lecture Notes in Computer
  Science}, pages 202--210. Springer, Cham.

\bibitem[{Liu et~al.(2026)Liu, Chen, and Jabbarvand}]{liu2024codemind}
Changshu Liu, Yang Chen, and Reyhaneh Jabbarvand. 2026.
\newblock \href {https://doi.org/10.1109/TSE.2026.3691250} {{CodeMind}:
  Evaluating large language models for implicit and explicit code execution
  reasoning}.
\newblock \emph{IEEE Transactions on Software Engineering}, 52(7):2031--2046.

\bibitem[{Liu et~al.(2023)Liu, Xia, Wang, and Zhang}]{liu2023evalplus}
Jiawei Liu, Chunqiu~Steven Xia, Yuyao Wang, and Lingming Zhang. 2023.
\newblock \href
  {https://proceedings.neurips.cc/paper_files/paper/2023/hash/43e9d647ccd3e4b7b5baab53f0368686-Abstract-Conference.html}
  {Is your code generated by {ChatGPT} really correct? {R}igorous evaluation of
  large language models for code generation}.
\newblock In \emph{Advances in Neural Information Processing Systems},
  volume~36, pages 21558--21572.

\bibitem[{Madeyski et~al.(2014)Madeyski, Orzeszyna, Torkar, and
  Jozala}]{madeyski2014slr}
Lech Madeyski, Wojciech Orzeszyna, Richard Torkar, and Mariusz Jozala. 2014.
\newblock \href {https://doi.org/10.1109/TSE.2013.44} {Overcoming the
  equivalent mutant problem: A systematic literature review and a comparative
  experiment of second order mutation}.
\newblock \emph{IEEE Transactions on Software Engineering}, 40(1):23--42.

\bibitem[{Maveli et~al.(2025)Maveli, Vergari, and Cohen}]{maveli2025seqcobench}
Nickil Maveli, Antonio Vergari, and Shay~B. Cohen. 2025.
\newblock \href {https://doi.org/10.18653/v1/2025.findings-naacl.382} {What can
  large language models capture about code functional equivalence?}
\newblock In \emph{Findings of the Association for Computational Linguistics:
  {NAACL} 2025}, pages 6880--6918. Association for Computational Linguistics.

\bibitem[{{Mistral AI}(2025)}]{mistral2025small3}
{Mistral AI}. 2025.
\newblock \href {https://mistral.ai/news/mistral-small-3} {{Mistral Small 3}}.
\newblock \url{https://mistral.ai/news/mistral-small-3}.

\bibitem[{Naeem et~al.(2020)Naeem, Lin, Naeem, and Liu}]{naeem2020ml}
Muhammad~Rashid Naeem, Tao Lin, Hamad Naeem, and Hailu Liu. 2020.
\newblock \href {https://doi.org/10.1002/smr.2238} {A machine learning approach
  for classification of equivalent mutants}.
\newblock \emph{Journal of Software: Evolution and Process}, 32(5):e2238.

\bibitem[{Necula(2000)}]{necula2000translation}
George~C. Necula. 2000.
\newblock \href {https://doi.org/10.1145/349299.349314} {Translation validation
  for an optimizing compiler}.
\newblock In \emph{Proceedings of the ACM SIGPLAN 2000 conference on
  Programming language design and implementation}, pages 83--94.

\bibitem[{Nica and Wotawa(2012)}]{nica2012constraints}
Simona Nica and Franz Wotawa. 2012.
\newblock \href {https://doi.org/10.4204/EPTCS.86.1} {Using constraints for
  equivalent mutant detection}.
\newblock In \emph{Proceedings of the Seventh Workshop on Model-Based Testing
  ({MBT} 2012)}, volume~86 of \emph{Electronic Proceedings in Theoretical
  Computer Science}, pages 1--8.

\bibitem[{Offutt and Craft(1994)}]{offutt1994compiler}
A.~Jefferson Offutt and W.~Michael Craft. 1994.
\newblock \href {https://doi.org/10.1002/stvr.4370040303} {Using compiler
  optimization techniques to detect equivalent mutants}.
\newblock \emph{Software Testing, Verification and Reliability}, 4(3):131--154.

\bibitem[{{OpenAI}(2025{\natexlab{a}})}]{openai2025gptoss}
{OpenAI}. 2025{\natexlab{a}}.
\newblock \href {https://doi.org/10.48550/arXiv.2508.10925} {gpt-oss-120b \&
  gpt-oss-20b model card}.
\newblock \emph{arXiv preprint arXiv:2508.10925}.

\bibitem[{{OpenAI}(2025{\natexlab{b}})}]{openai2025gpt41}
{OpenAI}. 2025{\natexlab{b}}.
\newblock \href {https://openai.com/index/gpt-4-1/} {Introducing {GPT-4.1} in
  the {API}}.
\newblock \url{https://openai.com/index/gpt-4-1/}.

\bibitem[{Oza et~al.(2026)Oza, Govil, Gupta, Khandelwal, Garg, and
  Singla}]{oza2025cetbench}
Neeva Oza, Ishaan Govil, Parul Gupta, Dinesh Khandelwal, Dinesh Garg, and Parag
  Singla. 2026.
\newblock \href {https://doi.org/10.18653/v1/2026.findings-acl.2070} {{LLMs}
  are brittle to simple code transformations: Introducing {CETBench} -- a
  benchmark for code-equivalence checking}.
\newblock In \emph{Findings of the Association for Computational Linguistics:
  ACL 2026}, pages 41653--41685. Association for Computational Linguistics.

\bibitem[{Papadakis et~al.(2015)Papadakis, Jia, Harman, and
  Le~Traon}]{papadakis2015tce}
Mike Papadakis, Yue Jia, Mark Harman, and Yves Le~Traon. 2015.
\newblock \href {https://doi.org/10.1109/ICSE.2015.103} {Trivial compiler
  equivalence: A large scale empirical study of a simple, fast and effective
  equivalent mutant detection technique}.
\newblock In \emph{2015 IEEE/ACM 37th IEEE International Conference on Software
  Engineering}, volume~1, pages 936--946. IEEE.

\bibitem[{Papadakis et~al.(2019)Papadakis, Kintis, Zhang, Jia, Le~Traon, and
  Harman}]{papadakis2019mutation}
Mike Papadakis, Marinos Kintis, Jie Zhang, Yue Jia, Yves Le~Traon, and Mark
  Harman. 2019.
\newblock \href {https://doi.org/10.1016/bs.adcom.2018.03.015} {Mutation
  testing advances: An analysis and survey}.
\newblock In \emph{Advances in Computers}, volume 112, pages 275--378.
  Elsevier.

\bibitem[{Peacock et~al.(2021)Peacock, Deng, Dehlinger, and
  Chakraborty}]{peacock2021astnn}
Samuel Peacock, Lin Deng, Josh Dehlinger, and Suranjan Chakraborty. 2021.
\newblock \href {https://doi.org/10.1109/ICSTW52544.2021.00016} {Automatic
  equivalent mutants classification using abstract syntax tree neural
  networks}.
\newblock In \emph{2021 IEEE International Conference on Software Testing,
  Verification and Validation Workshops (ICSTW)}, pages 13--18. IEEE.

\bibitem[{Person et~al.(2008)Person, Dwyer, Elbaum, and
  P{\u{a}}s{\u{a}}reanu}]{person2008differential}
Suzette Person, Matthew~B. Dwyer, Sebastian Elbaum, and Corina~S.
  P{\u{a}}s{\u{a}}reanu. 2008.
\newblock \href {https://doi.org/10.1145/1453101.1453131} {Differential
  symbolic execution}.
\newblock In \emph{Proceedings of the 16th ACM SIGSOFT International Symposium
  on Foundations of software engineering}, pages 226--237.

\bibitem[{Pnueli et~al.(1998)Pnueli, Siegel, and
  Singerman}]{pnueli1998translation}
Amir Pnueli, Michael Siegel, and Eli Singerman. 1998.
\newblock \href {https://doi.org/10.1007/BFb0054170} {Translation validation}.
\newblock In \emph{International Conference on Tools and Algorithms for the
  Construction and Analysis of Systems}, volume 1384 of \emph{Lecture Notes in
  Computer Science}, pages 151--166. Springer.

\bibitem[{Tian et~al.(2024)Tian, Shu, Wang, Cao, Kamei, and Chen}]{tian2024llm}
Zhao Tian, Honglin Shu, Dong Wang, Xuejie Cao, Yasutaka Kamei, and Junjie Chen.
  2024.
\newblock \href {https://doi.org/10.1145/3650212.3680395} {Large language
  models for equivalent mutant detection: How far are we?}
\newblock In \emph{Proceedings of the 33rd ACM SIGSOFT international symposium
  on software testing and analysis}, pages 1733--1745.

\bibitem[{van Hijfte and Oprescu(2021)}]{vanhijfte2021mutantbench}
Lars van Hijfte and Ana Oprescu. 2021.
\newblock \href {https://doi.org/10.1109/ICSTW52544.2021.00015} {{MutantBench}:
  An equivalent mutant problem comparison framework}.
\newblock In \emph{2021 IEEE International Conference on Software Testing,
  Verification and Validation Workshops (ICSTW)}, pages 7--12. IEEE.

\bibitem[{Wang et~al.(2023)Wang, Li, Qian, Yang, Wang, Shang, Kumar, Tan, Ray,
  Bhatia et~al.}]{wang2023recode}
Shiqi Wang, Zheng Li, Haifeng Qian, Chenghao Yang, Zijian Wang, Mingyue Shang,
  Varun Kumar, Samson Tan, Baishakhi Ray, Parminder Bhatia, et~al. 2023.
\newblock \href {https://doi.org/10.18653/v1/2023.acl-long.773} {{ReCode}:
  Robustness evaluation of code generation models}.
\newblock In \emph{Proceedings of the 61st Annual Meeting of the Association
  for Computational Linguistics (Volume 1: Long Papers)}, pages 13818--13843.
  Association for Computational Linguistics.

\bibitem[{Wei et~al.(2025)Wei, Cao, Li, Chen, Zhang, Wang, Liu, Teixeira, Yang,
  Wang, and Aiken}]{wei2025equibench}
Anjiang Wei, Jiannan Cao, Ran Li, Hongyu Chen, Yuhui Zhang, Ziheng Wang, Yuan
  Liu, Thiago S. F.~X. Teixeira, Diyi Yang, Ke~Wang, and Alex Aiken. 2025.
\newblock \href {https://doi.org/10.18653/v1/2025.emnlp-main.1718}
  {{EquiBench}: Benchmarking large language models' reasoning about program
  semantics via equivalence checking}.
\newblock In \emph{Proceedings of the 2025 Conference on Empirical Methods in
  Natural Language Processing}, pages 33868--33881. Association for
  Computational Linguistics.

\bibitem[{Yang et~al.(2025)Yang, Li, Yang, Zhang, Hui et~al.}]{yang2025qwen3}
An~Yang, Anfeng Li, Baosong Yang, Beichen Zhang, Binyuan Hui, et~al. 2025.
\newblock \href {https://doi.org/10.48550/arXiv.2505.09388} {Qwen3 technical
  report}.
\newblock \emph{arXiv preprint arXiv:2505.09388}.

\end{thebibliography}

\appendix
\section{Data generation details}
\label{app:datagen-details}

This appendix gives the details of the data generation in Section~\ref{sec:datagen}, from the choice of transformations to the stratified subset.

\subsection{Choice of transformations}
\label{app:transformations-choice}

We adapt the catalog from \textsc{SeqCoBench} \citep{maveli2025seqcobench}, which keeps our results comparable with prior code-equivalence work.
We omit its Identity-Operator-Misuse category, which changes only 3 of the 418 base programs, and its Dissimilar-Code-Selection category, which does not test reasoning about individual transformations.

The preserving transformations follow common practice.
NatGen uses variable renaming, dead-code insertion, operand swapping, and for-to-while conversion \citep{chakraborty2022natgen}, and RV and DCI\_V2 follow its \texttt{VAR\_i} naming and always-false guards.
AOM, LOM, and COM are restricted forms of the AOR, LCR, and ROR mutation operators \citep{papadakis2019mutation}, a standard way to inject bugs in software testing; they cover common operator mistakes, not all kinds of real-world bugs.

\subsection{Single-transformation examples}
\label{app:transformations}

This subsection illustrates each transformation in Table~\ref{tab:transformations} with a program from the base corpus.
Each example applies the transformation once to an original program.

AOM, LOM, COM, and OS change every matching location in the program and LT rewrites every \texttt{range} loop not nested in another \texttt{for} loop, whereas BLM, RV, DCI\_V1, and DCI\_V2 make a single edit.
Within a transformation sequence, each transformation applies to the current program, including code inserted by earlier transformations.

\paragraph{AOM (id 1, altering).}
Swaps \texttt{+} with \texttt{-} and \texttt{*} with \texttt{/}.
In the example, all three operators change, so the function no longer computes hexagonal numbers.

{\footnotesize Before}
\begin{promptbox}
def hexagonal_num(n):
    return n * (2 * n - 1)
\end{promptbox}
{\footnotesize After}
\begin{promptbox}
def hexagonal_num(n):
    return n / (2 / n + 1)
\end{promptbox}

\paragraph{BLM (id 2, altering).}
Flips the first \texttt{True} to \texttt{False} or, if the program contains no \texttt{True}, the first \texttt{False} to \texttt{True}.
In the example, the function now returns \texttt{False} for every input.

{\footnotesize Before}
\begin{promptbox}
def dif_Square(n):
    if n % 4 != 2:
        return True
    return False
\end{promptbox}
{\footnotesize After}
\begin{promptbox}
def dif_Square(n):
    if n % 4 != 2:
        return False
    return False
\end{promptbox}

\paragraph{LOM (id 3, altering).}
Swaps \texttt{and} with \texttt{or}.
In the example, the base case can no longer be reached, so every call recurses until Python raises a \texttt{RecursionError}.

{\footnotesize Before}
\begin{promptbox}
def newman_prime(n):
    if n == 0 or n == 1:
        return 1
    return 2 * newman_prime(n - 1) + newman_prime(n - 2)
\end{promptbox}
{\footnotesize After}
\begin{promptbox}
def newman_prime(n):
    if n == 0 and n == 1:
        return 1
    return 2 * newman_prime(n - 1) + newman_prime(n - 2)
\end{promptbox}

\paragraph{COM (id 4, altering).}
Swaps the leading operator of each comparison: \texttt{<} and \texttt{>}, \texttt{<=} and \texttt{>=}, and \texttt{==} and \texttt{!=}.
In the example, the sign test is reversed.

{\footnotesize Before}
\begin{promptbox}
def opposite_Signs(x, y):
    return x ^ y < 0
\end{promptbox}
{\footnotesize After}
\begin{promptbox}
def opposite_Signs(x, y):
    return x ^ y > 0
\end{promptbox}

\paragraph{RV (id 5, preserving).}
Renames the most frequently used variable or parameter to \texttt{VAR\_0}.
In the example, \texttt{XOR} is renamed.

{\footnotesize Before}
\begin{promptbox}
def search(arr):
    n = len(arr)
    XOR = 0
    for i in range(n):
        XOR = XOR ^ arr[i]
    return XOR
\end{promptbox}
{\footnotesize After}
\begin{promptbox}
def search(arr):
    n = len(arr)
    VAR_0 = 0
    for i in range(n):
        VAR_0 = VAR_0 ^ arr[i]
    return VAR_0
\end{promptbox}

\paragraph{DCI\_V1 (id 6, preserving).}
Inserts, at the start of the most deeply nested block, an assignment of a constant expression to a new variable that is never read.
The variable names come from a fixed vocabulary; during preprocessing we verified that none of them occurs in any original program.
In the example, the function body is the most deeply nested block, so the assignment to \texttt{window\_scope} is inserted before the \texttt{return}.

{\footnotesize Before}
\begin{promptbox}
def count(lst):
    return sum(lst)
\end{promptbox}
{\footnotesize After}
\begin{promptbox}
def count(lst):
    window_scope = 2 ** 20 >> 10 ^ 42, 'ab'[0] + chr(48), (3, 5, 7)[1]
    return sum(lst)
\end{promptbox}

\paragraph{DCI\_V2 (id 7, preserving).}
Inserts, at the same location as DCI\_V1, an \texttt{if} block whose condition is always false (the left-hand side evaluates to $31 \oplus 27 = 4$).
The block contains the same kind of unused assignment and an empty \texttt{try}/\texttt{except}.
In the example, the block is inserted before the \texttt{return}, where DCI\_V1 inserts its assignment.

{\footnotesize Before}
\begin{promptbox}
def count(lst):
    return sum(lst)
\end{promptbox}
{\footnotesize After}
\begin{promptbox}
def count(lst):
    if sum((13, 7, 11)) ^ 27 == 0:
        window_scope = 2 ** 20 >> 10 ^ 42, 'ab'[0] + chr(48), (3, 5, 7)[1]
        try:
            pass
        except:
            pass
    return sum(lst)
\end{promptbox}

\paragraph{OS (id 8, preserving).}
Swaps the operands of each \texttt{<}, \texttt{<=}, \texttt{>}, or \texttt{>=} comparison between two variables and mirrors the operator.
In the example, \texttt{a >= b} becomes \texttt{b <= a}.

{\footnotesize Before}
\begin{promptbox}
def maximum(a, b):
    if a >= b:
        return a
    else:
        return b
\end{promptbox}
{\footnotesize After}
\begin{promptbox}
def maximum(a, b):
    if b <= a:
        return a
    else:
        return b
\end{promptbox}

\paragraph{LT (id 9, preserving).}
Rewrites a \texttt{for} loop over \texttt{range} as an equivalent \texttt{while} loop with an explicit counter.
In the example, the counter \texttt{i} is initialized to $0$ before the loop and incremented at the end of its body.

{\footnotesize Before}
\begin{promptbox}
def search(arr):
    n = len(arr)
    XOR = 0
    for i in range(n):
        XOR = XOR ^ arr[i]
    return XOR
\end{promptbox}
{\footnotesize After}
\begin{promptbox}
def search(arr):
    n = len(arr)
    XOR = 0
    i = 0
    while i < n:
        XOR = XOR ^ arr[i]
        i += 1
    return XOR
\end{promptbox}

\subsection{Transformation-sequence example}
\label{app:example-pair}

This subsection shows a pair $(p, p')$ from the stratified subset together with its transformation sequence set.
The original program $p$ is:

\begin{promptbox}
def check_smaller(test_tup1, test_tup2):
    return all(x > y for x, y in zip(test_tup1, test_tup2))
\end{promptbox}

The modified program $p'$ is:

\begin{promptbox}
def check_smaller(test_tup1, test_tup2):
    if sum((13, 7, 11)) ^ 27 == 0:
        cursor_scale = 2 ** 20 >> 10 ^ 42, 'ab'[0] + chr(48), (3, 5, 7)[1]
        try:
            pass
        except:
            pass
    return all(y > x for x, y in zip(test_tup1, test_tup2))
\end{promptbox}

Its transformation sequence set is $\Pi(p, p') = \{(4, 8, 7), (8, 4, 7)\}$, so $k(p, p') = 3$.
The pair is non-equivalent ($y{=}0$) because COM (id~4) is altering.
COM turns \texttt{x > y} into \texttt{x < y}, and OS (id~8) rewrites \texttt{x < y} as \texttt{y > x}.
Applied in the other order, OS and COM produce the same comparison, so both sequences belong to $\Pi(p, p')$.
In T3, both sequences count as correct, and in T2 the correct answer is id~4.

The position of DCI\_V2 (id~7) matters for two reasons.
First, if DCI\_V2 comes before COM, COM also flips the comparison \texttt{== 0} in the inserted \texttt{if} block to \texttt{!= 0}.
Second, DCI\_V2 chooses the name of the inserted variable from a hash of the program it modifies, so the sequence $(4, 7, 8)$ inserts \texttt{token\_depth} instead of \texttt{cursor\_scale}.
The other orders therefore produce three further modified programs, each forming a separate pair in \dataset{}, with transformation sequence sets $\{(4, 7, 8)\}$, $\{(7, 4, 8), (7, 8, 4)\}$, and $\{(8, 7, 4)\}$.

\subsection{Generation procedure}
\label{app:algorithm}

$\textsc{Extend}$ (line~11 of Algorithm~\ref{alg:collections}) takes every valid prefix of the parent collection and appends every transformation of the pool whose id is not already in the prefix.
For each new prefix, it checks Definition~\ref{def:valid-transformation-sequence}: it compares the ASTs before and after the transformation and runs the tests in $E$ on the new program with a time limit of 30 seconds.
$\textsc{Extend}$ stores every new prefix together with whether it is valid, so the 20 collections also record invalid prefixes.
This allows an audit of the generation by the size identity $|D_{k,a}| = V(D_{\mathrm{parent}}) \cdot m$.
Here $V(\cdot)$ counts the valid prefixes of the parent collection, and $m$ is the number of pool transformations not yet in a prefix.
If the pool is $\mathcal{A}$, then $m{=}4$.
If the pool is $\mathcal{P}$ and the parent prefixes already contain $j$ preserving transformations, then $m{=}5-j$.
A transformation that crashes, or any other silently dropped prefix, violates this identity.

\subsection{Integrity checks}
\label{app:aggregate}

After the pairs are formed, the pipeline aborts if any of the following checks fails.
\begin{itemize}
\item[(i)] Every sequence in $\Pi(p, p')$ uses distinct transformation ids and contains at most one altering transformation.
\item[(ii)] All sequences in $\Pi(p, p')$ have the same length $k(p, p')$.
\item[(iii)] Either every sequence in $\Pi(p, p')$ contains an altering transformation or none does, and this matches the label $y$.
\item[(iv)] No pair is labeled both equivalent and non-equivalent.
\item[(v)] For each length $k$, the sets $\Pi(p, p')$ of all pairs together contain exactly the valid sequences generated at that length, so no sequence is lost or counted twice.
\end{itemize}
Checks~(ii) and~(iii) back the definitions of $k$ and $y$ (Definition~\ref{def:transformation-sequence-set} and Section~\ref{sec:datagen:corpus}).
No transformation in the catalog undoes another: renames are never reverted, inserted dead code is never removed, the \texttt{for}$\rightarrow$\texttt{while} rewrite has no inverse, and each operator swap, which is its own inverse, appears at most once in a sequence.
We therefore expect no two sequences of different lengths to produce the same pair, and check~(ii) confirms this for every pair.

\subsection{Corpus statistics}
\label{app:corpus-stats}

Table~\ref{tab:stats} breaks down the 67{,}658 labeled pairs of \dataset{} by transformation sequence length $k$ and label.
The non-equivalent pairs at $k{=}1$ form the smallest stratum and limit the number of pairs drawn per stratum (Section~\ref{sec:datagen:corpus}).

\begin{table}[t]
\caption{Number of labeled pairs by transformation sequence length $k$.}
\label{tab:stats}
\begin{center}
\small
\setlength{\tabcolsep}{4pt}
\begin{tabular}{@{}lrrr@{}}
\toprule
\textbf{$k$} & \textbf{Equivalent pairs} & \textbf{Non-equivalent pairs} & \textbf{Total} \\
\midrule
1     & 1{,}315  & 464     & 1{,}779  \\
2     & 2{,}958  & 2{,}818  & 5{,}776  \\
3     & 4{,}098  & 9{,}889  & 13{,}987 \\
4     & 2{,}760  & 20{,}367 & 23{,}127 \\
5     & 1{,}379  & 21{,}610 & 22{,}989 \\
\midrule
Total & 12{,}510 & 55{,}148 & 67{,}658 \\
\bottomrule
\end{tabular}
\end{center}
\end{table}

\subsection{Stratified sampling}
\label{app:sampling}

The stratified subset draws $n = 450$ pairs from each of the ten strata, one for each combination of $k$ and $y$, for 4{,}500 pairs in total.

\paragraph{Coverage target.}
The \emph{presence set} of a pair is the set of transformation ids that occur in at least one sequence in $\Pi(p, p')$.
For each stratum, $R(k, y)$ is the set of ids whose coverage the sampler maximizes.
For a set $S$ of transformations, $\mathrm{ids}(S)$ is the set of their ids.
For the equivalent strata ($y{=}1$), $R = \mathrm{ids}(\mathcal{P})$, because no altering id occurs there.
For the stratum $(k{=}1, y{=}0)$, $R = \mathrm{ids}(\mathcal{A})$, because each sequence there is a single altering transformation.
For the other non-equivalent strata, $R = \mathrm{ids}(\mathcal{T})$.

\paragraph{Maximin cover.}
The sampler looks for the largest threshold $x$ such that some $n$ pairs of the stratum contain every id of $R$ in the presence sets of at least $x$ of them.
It binary-searches over $x$ and tests each threshold with a randomized greedy set-multicover heuristic, which repeatedly adds the pair that covers the most ids still below the threshold.
The heuristic makes up to three attempts per threshold.
The sampler keeps the cover found for the largest feasible $x$, fills the remaining slots with pairs drawn uniformly at random without replacement, and shuffles the result.

\paragraph{Seeds and T2 subset.}
All draws are deterministic: each stratum's random generator is seeded with a CRC-32 hash of a descriptor that names the stratum and includes a global seed (42), and the final shuffle uses the global seed.
The T2 subset consists of the 2{,}250 non-equivalent pairs of the same draw (450 per $k$).
It is obtained by filtering, not by resampling, so T2 uses the same non-equivalent pairs as T1 and T3.

\paragraph{Verification.}
A final pass checks the size of every stratum, the coverage threshold achieved in each, and that no equivalent stratum contains an altering id.

\section{Evaluation prompts}
\label{app:prompts}

This appendix gives the prompt templates for the three tasks and the rule that maps each response to a prediction.
We send each prompt as a single user message with no system prompt.
We reproduce the templates verbatim, including the Markdown markup that the model receives, such as \texttt{**...**}.
The symbol \ensuremath{\hookrightarrow} marks a line wrapped for display only.
In every template, the original program $p$ replaces \texttt{\{code\_1\}}, and the modified program $p'$ replaces \texttt{\{code\_2\}}.
In the T2 and T3 templates, the transformation catalog of Appendix~\ref{app:prompts:catalog} replaces \texttt{\{transformation\_description\}}.

\paragraph{Prompt variants.}
Each variant fills two slots of a shared template (Table~\ref{tab:prompt-variants}).
The \emph{answer instruction} either asks for the answer alone (direct) or asks the model to reason step by step and write the answer on the last line (CoT).
The \emph{example} is either a worked one-shot example --- a program pair together with its correct answer --- or a format-only example that shows the expected answer line without a program pair.
The one-shot example for T1 and T2 is a single altering transformation (id~1, AOM) applied to an MBPP program; the one for T3 applies transformations 5, 7, and 1 in that order.
For each task, Appendices~\ref{app:prompts:rq1}--\ref{app:prompts:rq3} give the full \pcotshot{} prompt used for the model comparisons in Section~\ref{sec:exp:setup}, followed by the text that the other three variants replace.

\begin{table}[t]
\caption{Composition of the four prompt variants.}
\label{tab:prompt-variants}
\begin{center}
\small
\begin{tabular}{@{}lll@{}}
\toprule
\textbf{Variant} & \textbf{Answer instruction} & \textbf{Example} \\
\midrule
\pbase{}    & direct & format-only$^{\dagger}$ \\
\pcot{}     & CoT    & format-only             \\
\pshot{}    & direct & one-shot                \\
\pcotshot{} & CoT    & one-shot                \\
\bottomrule
\end{tabular}
\end{center}
{\footnotesize $^{\dagger}$The T1 \pbase{} prompt has no example.}
\end{table}

\subsection{Transformation catalog}
\label{app:prompts:catalog}

The T2 and T3 prompts list all nine transformations of Table~\ref{tab:transformations}, one per line, as the id followed by the operational specification below.
The T2 prompts list all nine, not only the four altering transformations, so the model must itself determine which transformation altered the semantics.

\begin{promptbox}
1: Scope: binary arithmetic operators (+, -, *, /) and augmented assignments (+=, -=, *=, /=) only; does not modify comparisons, logical operators, boolean literals, or variable names. Procedure: parse to AST; visit every BinOp and AugAssign node; replace each operator with its paired opposite (+↔-, -↔+, *↔/, /↔*); unparses AST back to source. Example: `x = a + b` → `x = a - b`; `z += 10` → `z -= 10`.
2: Scope: boolean literals `True` and `False` in source text only; at most one replacement per application (first match in file order). Does not change `and`/`or`, comparisons, or non-literal booleans. Procedure: scan source as text; if `True` appears, replace the first `True` with `False`; else if `False` appears, replace the first `False` with `True`; otherwise leave code unchanged. Example: `if True: return 1` → `if False: return 1`.
3: Scope: logical boolean operators in BoolOp nodes (`and`, `or`) only; does not change arithmetic, comparisons, or `True`/`False` literals. Procedure: parse to AST; visit each BoolOp; swap `and`↔`or` on that node; unparses AST back to source. Example: `if x > 0 and x < 10:` → `if x > 0 or x < 10:`.
4: Scope: comparison operators in Compare nodes: >, <, >=, <=, ==, != (first operator in a compare chain only); does not swap operands or change `in`/`is` tests. Procedure: parse to AST; visit Compare nodes; map each operator to its opposite (>↔<, >=↔<=, ==↔!=); unparses AST back to source. Example: `if a == b:` → `if a != b:`; `while n > 0:` → `while n < 0:`.
5: Scope: user-defined variable and parameter identifiers; renames the single most frequent name (default top_n=1) to generated names VAR_0, VAR_1, … that do not appear in the original source. Procedure: parse to AST; count Name and arg occurrences; pick the most common identifier(s); rewrite every load/store/del and matching parameter to VAR_i; unparses AST back to source. Example: `result = data + offset` → `VAR_0 = data + offset` when `result` is the most common name.
6: Scope: inserts one new statement into the deepest statement-owning block (module, function, class, if/for/while/try, etc.); no existing lines are removed or altered. Procedure: parse to AST; locate the deepest `body` list; insert a standalone Assign to a plausible unused name (e.g. cache_limit) with an opaque constant tuple expression (bit ops, chr, indexing); the name is formed as `{word_a}_{word_b}` from fixed vocabularies and is guaranteed not to appear anywhere in the source before this transformation runs (no collision with existing identifiers); position at middle of module body or start of nested blocks; unparses AST back to source. Example: adds `cache_limit = ((2 ** 20 >> 10) ^ 0x2A), ...` inside the deepest block.
7: Scope: same deepest-block targeting as DCI_V1; adds one never-taken `if` block without modifying reachable code. Procedure: parse to AST; find deepest statement block; insert `if` with a statically false condition (e.g. `(sum((13, 7, 11)) ^ 27) == 0)`); body holds an unused assign plus an inert try/except with only `pass`; the dead assign uses the same fresh-name rule as DCI_V1 (guaranteed absent from the original code before transformation); same insert position rules as DCI_V1; unparses AST back to source. Example: adds `if (sum((13, 7, 11)) ^ 27) == 0: token_scale = ...; try: pass except: pass` inside the deepest block.
8: Scope: Compare nodes where both operands are simple names (ast.Name); operators >, <, >=, <= only (equality == and != are skipped). Procedure: parse to AST; for each qualifying comparison, swap left and right operands and replace the operator with the equivalent flipped form (>↔<, >=↔<=); unparses AST back to source. Example: `if a > b:` → `if b < a:`; `while low <= high:` → `while high >= low:`.
9: Scope: `for` loops whose iterator is a direct `range(...)` call (one, two, or three arguments); other loop forms are left unchanged. Procedure: parse to AST; for each matching For node, emit (1) an assignment initializing the loop variable to the range start, (2) a While with end test (< for positive step, > for negative step), (3) a body equal to the original loop body plus an increment/decrement by the step; unparses AST back to source. Example: `for i in range(3): body` → `i = 0` then `while i < 3: body; i += 1`.
\end{promptbox}

\subsection{T1: Detecting semantic equivalence}
\label{app:prompts:rq1}

\paragraph{\pcotshot{} prompt.}

\begin{promptbox}
You are a Python code expert.
Determine if the following Code1 and Code2 are functionally equivalent, meaning that they have the same output for any given input.

Think step by step: compare Code1 and Code2, identify any differences in logic, control flow, edge-case handling, or return values, and reason about whether those differences could produce different outputs for any input.
After your analysis, write the final answer on the last line as exactly "YES" or "NO", with no other text on that line.


---
**Example:**

Input:

Code1:
def first_repeated_char(str1):
    for index, c in enumerate(str1):
        if str1[:index + 1].count(c) > 1:
            return c


Code2:
def first_repeated_char(str1):
    for index, c in enumerate(str1):
        if str1[:index - 1].count(c) > 1:
            return c


The last line of the output:
NO


---
**Task:**
Run the task on the following input:


Code1:
{code_1}


Code2:
{code_2}
\end{promptbox}

\paragraph{Other variants.}
\pbase{} and \pshot{} replace the two-line CoT instruction (``Think step by step: \ldots'') with the direct instruction:

\begin{promptbox}
Respond with exactly "YES" if they are equivalent or "NO" if they are not.
Do not include any additional text or explanation.
\end{promptbox}

\pshot{} keeps the one-shot example and labels its answer \texttt{Output:} instead of \texttt{The last line of the output:}.
\pbase{} and \pcot{} remove everything from the first \texttt{-{}-{}-} line through \texttt{Run the task on the following input:}.
In its place, \pcot{} shows the format-only example \texttt{**Example of the last line of the response:**} followed by \texttt{NO}; \pbase{} shows no example.

\paragraph{Parser.}
The parser takes the last non-empty line of the response, strips surrounding whitespace, and accepts it only if it equals \texttt{YES} or \texttt{NO}, ignoring case.
Any other response is unparseable.

\subsection{T2: Diagnosing semantic change}
\label{app:prompts:rq2}

\paragraph{\pcotshot{} prompt.}

\begin{promptbox}
You are a Python code expert.
You are given two code snippets **Code1** and **Code2**.
**Code 2** is produced by applying one or more transformations on **Code1**.
Exactly one of the applied transformations alters the semantic of the code snippet, while other applied transformations (if any) preserve the semantic.
Your task is to determine which transformation was applied to **Code 1** to alter its semantics and produce **Code 2**.

Think step by step: compare Code 1 and Code 2, list the syntactic differences, and identify which transformation ID best explains the semantic change. After your analysis, write the final answer on the last line as a single digit only (one of 1, 2, 3, 4, 5, 6, 7, 8, 9), with no other text on that line.

---
Below is a list of all possible transformations, each with an ID and description.
**Available Transformations:**
{transformation_description}

---
**Example:**

Input:

**Code 1:**
def first_repeated_char(str1):
    for index, c in enumerate(str1):
        if str1[:index + 1].count(c) > 1:
            return c

**Code 2:**
def first_repeated_char(str1):
    for index, c in enumerate(str1):
        if str1[:index - 1].count(c) > 1:
            return c


The last line of the output:
1

---
**Task:**
Run the task on the following input:

**Code 1:**
{code_1}

**Code 2:**
{code_2}
\end{promptbox}

\paragraph{Other variants.}
\pbase{} and \pshot{} replace the CoT instruction (``Think step by step: \ldots'') with the direct instruction:

\begin{promptbox}
Respond with exactly one digit and nothing else. Your answer must be exactly one of: 1, 2, 3, 4, 5, 6, 7, 8, 9.
\end{promptbox}

\pshot{} keeps the one-shot example and labels its answer \texttt{Output:} instead of \texttt{The last line of the output:}.
\pbase{} and \pcot{} remove the one-shot example, from \texttt{**Example:**} through \texttt{Run the task on the following input:}.
Instead, they show a format-only example directly after the answer instruction, before the catalog.
They also omit the \texttt{-{}-{}-} line above the catalog and add one between \texttt{**Code 1:**} and \texttt{**Code 2:**}.
In \pbase{}, this example is \texttt{**Example Response:**} followed by \texttt{4}.
In \pcot{}, it is \texttt{**Example of the last line of the response:**} followed by \texttt{4}.

\paragraph{Parser.}
The parser extracts the first integer on the last non-empty line of the response; if that line contains none, it falls back to the first integer anywhere in the response.
The response is unparseable if no integer is found or the integer is not in $\mathrm{ids}(\mathcal{T}) = \{1,\dots,9\}$.
We count a parseable prediction of a preserving id (5--9) as incorrect.

\subsection{T3: Reconstructing transformation sequences}
\label{app:prompts:rq3}

\paragraph{\pcotshot{} prompt.}

\begin{promptbox}
You are a Python code expert.
Your task is to determine which transformations were applied (in order) to **Code 1** to produce **Code 2**.

Think step by step: compare Code 1 and Code 2 carefully, identify every syntactic or semantic change, match each change to the most appropriate transformation ID from the available transformations, and determine the order in which the transformations must have been applied.
After your analysis, write the final answer on the last line as only a comma-separated list of transformation IDs in the order they were applied, with no other text on that line.

Below is a list of all possible transformations, each with an ID and description.
**Available Transformations:**
{transformation_description}

---
**Example:**

**Code 1:**
def div_list(nums1, nums2):
    result = map(lambda x, y: x / y, nums1, nums2)
    return list(result)

**Code 2:**
def div_list(VAR_0, nums2):
    if sum((13, 7, 11)) ^ 27 == 0:
        salt_state = 2 ** 20 >> 10 ^ 42, 'ab'[0] - chr(48), (3, 5, 7)[1]
        try:
            pass
        except:
            pass
    result = map(lambda x, y: x * y, VAR_0, nums2)
    return list(result)


**Output:**
5, 7, 1


---
**Task:**
Run the task on the following input:

---

**Code 1:**
{code_1}

---

**Code 2:**
{code_2}
\end{promptbox}

\paragraph{Other variants.}
\pbase{} and \pshot{} replace the two-line CoT instruction (``Think step by step: \ldots'') with the direct instruction:

\begin{promptbox}
Please respond **only** with a comma-separated list of transformation IDs in the order they were applied.
Do not include explanations or any additional text.
\end{promptbox}

\pshot{} keeps the one-shot example unchanged.
\pbase{} and \pcot{} remove the one-shot example, from the \texttt{-{}-{}-} line above \texttt{**Example:**} through \texttt{Run the task on the following input:}.
Instead, they show a format-only example directly after the answer instruction, before the catalog.
In \pbase{}, this example is \texttt{**Example Response:**} followed by \texttt{5, 7, 1}.
In \pcot{}, it is \texttt{**Example of the last line of the response:**} followed by \texttt{5, 7, 1}.
No variant discloses the transformation sequence length $k$.

\paragraph{Parser.}
The parser reads the final line of the response and splits it on commas.
From each comma-separated field it takes the first integer and keeps it if it is in $\mathrm{ids}(\mathcal{T})$, preserving order, and drops out-of-range integers.
The result is the predicted transformation sequence $\hat{\pi}$, which Section~\ref{sec:exp:rq3} counts as correct if it is one of the sequences in $\Pi(p, p')$.
If no integer is kept, the response is unparseable.

\section{Reproducibility}
\label{app:reproducibility}
Section~\ref{sec:datagen} and Appendix~\ref{app:datagen-details} specify the full data generation pipeline, including the generation algorithm (Appendix~\ref{app:algorithm}), the integrity checks on the resulting pairs (Appendix~\ref{app:aggregate}), and the stratified sampling procedure with its fixed seeds (Appendix~\ref{app:sampling}).
Appendix~\ref{app:prompts} gives every prompt template verbatim together with the response parsers.
Section~\ref{sec:exp:setup} gives the inference settings: temperature $0$, a $4{,}000$-token budget, and one response per pair.
We query all models through the OpenRouter API using the identifiers
\nolinkurl{qwen/qwen3-235b-a22b-2507}, \nolinkurl{openai/gpt-oss-120b}, \nolinkurl{mistralai/mistral-small-24b-instruct-2501}, \nolinkurl{meta-llama/llama-3.1-8b-instruct}, and \nolinkurl{openai/gpt-4.1-nano}.
The experiments issue 94{,}500 queries, not counting repeated queries.
Each task uses eight combinations of model and prompt variant, namely the four prompt variants on Qwen3 and \pcotshot{} on the other four models, and T1 additionally uses \pcot{} on GPT-4.1-nano.
Each combination covers 4{,}500 pairs in T1 and T3 and 2{,}250 pairs in T2.
We will publicly release the code for data generation and experiments, the full corpus of 67{,}658 pairs, and all raw model responses upon conference acceptance.
Providers can update hosted models over time, and OpenRouter can serve the same model from different providers.
Decoding at temperature $0$ is also not always deterministic.
Re-querying the models can therefore yield slightly different numbers.
The released responses will let readers recompute every reported number exactly.

\end{document}